\documentclass[aps,pre,reprint,longbibliography,superscriptaddress]{revtex4-2}
\usepackage[T1]{fontenc}
\usepackage{subcaption}
\usepackage{graphicx}
\usepackage{dcolumn}
\usepackage{bm}
\usepackage{amssymb}
\usepackage{amsmath}
\usepackage{mathrsfs}
\usepackage[colorlinks=true,linkcolor=blue,citecolor=blue,urlcolor=blue ]{hyperref}
\begin{document}

\title{Fast-erasure energy-cost lower bound and its first-order correction\texorpdfstring{\\}{ }under restricted control}

\author{Yihang Cai}
\thanks{These two authors contributed equally to this work.}
\affiliation{Department of Physics and Shenzhen Research Institute, Xiamen University, Xiamen 361005, Fujian, China}
\author{Yirong Jiang}
\thanks{These two authors contributed equally to this work.}
\affiliation{Department of Physics and Shenzhen Research Institute, Xiamen University, Xiamen 361005, Fujian, China}
\author{Chenlong Huang}
\affiliation{Department of Physics and Shenzhen Research Institute, Xiamen University, Xiamen 361005, Fujian, China}
\author{Yue Liu}
\affiliation{Center for Gravitational Physics and Quantum Information, Yukawa Institute for Theoretical Physics, Kyoto University, Kitashirakawa Oiwakecho, Sakyo-ku, Kyoto 606-8502, Japan}
\author{Dahai He}
\email[Contact author: ]{dhe@xmu.edu.cn}
\affiliation{Department of Physics and Shenzhen Research Institute, Xiamen University, Xiamen 361005, Fujian, China}

\date{November 13, 2026}

\begin{abstract}
\label{sec:abstract}
Determining the minimum work required for information erasure is central to understanding the energetic limits of fast and reliable information processing. In the fast-erasure limit, we derive an explicit expression for this minimum work in a one-dimensional overdamped isothermal system with control restricted to linear and quadratic additions to a reference potential. For symmetric initial distributions at fixed mobility, the minimum work scales inversely with the erasure duration, with a prefactor determined solely by the second moment and a quantile associated with the target success probability. We further derive the leading finite-time correction at a prescribed success probability and verify both analytical predictions numerically. These results provide an analytical benchmark for fast erasure under restricted control, connecting a universal leading cost with finite-time corrections that account for the detailed form of the memory potential.
\end{abstract}

\maketitle

\section{Introduction}
\label{sec:intro}

Landauer's principle states that complete erasure of a bit stored in two equivalent logical states with equal initial occupation probabilities requires an average work of at least $k_{\mathrm{B}}T\ln 2$, a bound approached in the quasistatic limit~\cite{Landauer1961,Bennett1982}. This bound has been verified in colloidal double-well and feedback-trap experiments and supported by numerical simulations of nanomagnets~\cite{Berut2012,Jun2014,Lambson2011}. In practice, however, storage devices must complete write--read--reset cycles in finite time. Shortening the erasure duration increases the cycling rate but generally leads to greater nonequilibrium dissipation~\cite{Dago2022,Boyd2022,Lee2022,Zhen2021}. Erasure protocols must also achieve the target success probability within the prescribed duration using the available control parameters~\cite{Li2024PRE,Lyu2026}. Minimizing the erasure work at a prescribed duration and success probability is therefore a fundamental problem in finite-time information thermodynamics~\cite{Proesmans2020PRL,Proesmans2020PRE,Ma2022,Li2024PRE}.

For overdamped systems, this optimization problem is closely related to optimal transport~\cite{Aurell2011,Aurell2012,Nakazato2021,Oikawa2025}. When the potential landscape can be manipulated arbitrarily, the minimum irreversible work required to connect fixed initial and final distributions is characterized by the $L^2$-Wasserstein distance~\cite{Aurell2011,Aurell2012}. Within this full-control setting, the optimal erasure work and its fast-erasure limit have also been obtained with only the final success probability prescribed~\cite{Proesmans2020PRE}. In this limit, the freedom in the final distribution allows the optimal protocol to reduce irreversible work by preferentially transporting probability mass near the logical-state boundary.

In realistic systems, however, control is usually limited to a few potential parameters, such as the barrier height and tilt in double-well information erasure~\cite{Saira2020,Zulkowski2014,Li2024PRE}. Optimal protocols under such restricted control have been studied in several paradigmatic models~\cite{Schmiedl2007,Then2008,Zulkowski2014,Plata2019PRE,Blaber2022EPL,Li2024PRE,Lyu2026}. In the slow-driving limit, linear-response theory expresses the excess work as a time integral of a quadratic form in the rates of change of the control parameters, so that optimal protocols correspond to geodesics of the thermodynamic metric~\cite{Sivak2012,SivakCrooks2016PRE}. In the opposite fast-driving limit, short-time expansions provide asymptotic predictions for the optimal work and protocols~\cite{Blaber2021}. Beyond these limiting regimes, optimal protocols have been obtained by applying Pontryagin's maximum principle to the Fokker--Planck dynamics and solving the resulting system of Hamiltonian partial differential equations~\cite{Zhong2022}.

These studies~\cite{Proesmans2020PRE,Zulkowski2014,Zhong2022,Li2024PRE} motivate a natural question: can the minimum work of fast erasure be determined analytically under restricted control when the final state is constrained only by a prescribed success probability? Here we show that linear--quadratic control leads to affine probability transport in the fast-erasure limit, reducing the work minimization to a convex optimization over translation and scaling. For symmetric initial distributions, this yields explicit optimal protocols and a minimum work that scales inversely with the erasure duration. At fixed mobility, the prefactor depends only on the initial second moment and the quantile associated with the target success probability. We further derive the leading finite-time correction, accounting for both the work correction under the leading-order protocol and the protocol adjustment required to maintain the prescribed success probability. The predictions are verified against the short-time expansion of the exact harmonic solution and numerical optimization in a quartic double-well potential.

The remainder of this paper is organized as follows. Section~\ref{sec:theory} develops the theoretical framework, covering the thermodynamic formulation, the minimum work in the fast-erasure limit, and the first-order finite-time correction. Section~\ref{sec:numerical} compares the short-time predictions with the exact harmonic solution and describes the numerical optimization for a quartic double-well potential. Section~\ref{sec:results} compares the analytical predictions with the numerical optima and examines deviations from the fast-erasure approximation as the erasure duration increases. Finally, Sec.~\ref{sec:conclusion} summarizes the main results and discusses possible extensions. Appendix~\ref{app:first-order} provides the detailed derivation of the first-order correction, while Appendix~\ref{app:numerical} describes the numerical implementation and verification.
\section{Theoretical framework}
\label{sec:theory}

\subsection{Model and thermodynamic formulation}
\label{subsec:model}
For a one-dimensional overdamped Brownian particle in contact with a single heat bath of temperature $T$, the dynamics is described by the overdamped Langevin equation
\begin{equation}
    \mathrm{d}X_t = -\mu\,\partial_{x}U(X_t,t)\,\mathrm{d}t + \sqrt{2D}\,\mathrm{d}\mathcal{B}_t,
\end{equation}
where $\mu$ is the mobility, $D=\mu k_{\mathrm{B}}T$ the diffusion constant, and $B_t$ a Wiener process. The probability density evolves according to the Fokker--Planck equation
\begin{equation}
    \partial_{t}\rho = -\partial_{x}J,
    \label{eq:continuous_equation}
\end{equation}
with
\begin{equation}
    J = -\mu\rho\,\partial_{x}U - D\,\partial_{x}\rho,
\end{equation}
and we introduce the probability-flux velocity
\begin{equation}
    v = \frac{J}{\rho} = -\mu\,\partial_{x}U - D\,\partial_{x}\ln\rho,
    \label{eq:volecity}
\end{equation}
which is useful for calculating the dissipated work.

For our erasure process, the initial state $\rho_0$ is the equilibrium distribution of a reference potential $U_{\mathrm{ref}}(x)$, which is an even function of position,
\begin{equation}
    \rho_0(x)=\frac{e^{-U_{\mathrm{ref}}(x)/(k_{\mathrm{B}}T)}}{Z_0}.
\end{equation}
We divide the whole $x$ axis into two parts, where $x<0$ is the left state and $x>0$ is the right state. From the evenness of $U_{\mathrm{ref}}$ it follows directly that the left and right states initially have equal probability, and that the mean position $\langle X_0\rangle_0$ vanishes. To ensure a finite leading fast-erasure coefficient, we require a finite second moment $m_2 = \int_{-\infty}^{+\infty}x^2\rho_0(x)\,dx$ and that $\rho_0$ be absolutely continuous.
We erase the bit to the right state; the success-rate constraint is
\begin{equation}
    P_{\mathrm{suc}}(\tau) = \int_{0}^{\infty}\rho(x,\tau)\,\mathrm{d}x \geq p,\quad \frac{1}{2}<p<1.
    \label{eq:success_rate}
\end{equation}
The total potential takes the form
\begin{equation}
    U(x,t)=U_{\mathrm{ref}}(x) + u_1(t)x + \frac{1}{2}u_2(t)x^2,
    \label{eq:control_potential}
\end{equation}
where $u_1$ and $u_2$ together can be interpreted as controlling an extra harmonic term; in the quartic bistable reference potential of Sec.~\ref{sec:114514}, $u_1$ produces a tilt while $u_2$ modifies the barrier height. We require that the potential equal $U_{\mathrm{ref}}$ both before and after the erasure; discontinuous jumps in the protocols are allowed.

To quantify the thermodynamic cost of erasure, we recall the work expression. Using Eq.~(\ref{eq:volecity}) and integration by parts, we arrive at the work--dissipation relation of stochastic energetics,
\begin{equation}
    \begin{aligned}
    W_{\mathrm{irr}} &= \langle W\rangle - \Delta\mathscr{F}_{\mathrm{neq}}\\ 
    &= \frac{1}{\mu}\int_{0}^{\tau}\mathrm{d}t\int_{-\infty}^{+\infty}\mathrm{d}x\,\rho(x,t)v^{2}(x,t).
    \end{aligned}
    \label{eq:irr_work}
\end{equation}
The nonequilibrium free energy is defined as
\begin{equation}
    \mathscr{F}_{\mathrm{neq}}[\rho,U]=\int_{-\infty}^{\infty}\rho(x)\bigl[U(x)+k_{\mathrm{B}}T\ln\rho(x)\bigr]\,\mathrm{d}x.
    \label{eq:non_free}
\end{equation}
From Eq.~(\ref{eq:irr_work}) we see that a quench of the potential contributes to the total work only through the change in the nonequilibrium free energy, while having no direct impact on the dissipated irreversible work.
Because the probability distribution must be transported by a finite amount during fast erasure, the probability-flux velocity in Eq.~(\ref{eq:volecity}) scales as $O(\tau^{-1})$; consequently the leading term of $W_{\mathrm{irr}}$ diverges as $1/\tau$. Besides, finite $\Delta\mathscr{F}_{\mathrm{neq}}$ is $O(1)$, so $\tau W_{\mathrm{irr}}$ and $\tau\langle W\rangle$ share the same zeroth-order term, while their first-order difference is $\tau\Delta\mathscr{F}_{\mathrm{neq}}$. Hence, the zeroth-order coefficient $C_0$ can be derived from $W_{\mathrm{irr}}$, while the first-order correction and its comparison with numerical results are phrased in terms of $\tau\langle W\rangle$.

\subsection{Fast-erasure limit and minimum dissipation}
\label{subsec:fast-erasure}
In this section we consider the optimization of Eq.~(\ref{eq:irr_work}) in the fast-erasure limit. We define the scaled
time $s=t/\tau \in [0,1]$, the rescaled probability density $\rho_{\tau}(x,s)=\rho(x,\tau s)$, and the probability-flux 
velocity $V_{\tau}(x,s)=\tau v(x,\tau s)$. Rewriting Eq.~(\ref{eq:continuous_equation}) and Eq.~(\ref{eq:irr_work}) gives
    \begin{equation}
        \partial_s \rho_{\tau} + \partial_x (\rho_{\tau}V_{\tau}) = 0,
        \label{eq:continuous_equation_rewrite}
    \end{equation}
    \begin{equation}
        \tau W_{\mathrm{irr}} = \frac{1}{\mu}\int_0^1 \mathrm{d} s \int_{-\infty}^{+\infty} \mathrm{d}x \,\rho_{\tau}V^2_{\tau}.
    \end{equation}
Because the time dependence is realized explicitly through $u_1(t)$ and $u_2(t)$, these protocols share the same scaling $O(\tau^{-1})$ as the probability-flux velocity; we rescale them as $\bar{u}_i(s)=\tau u_i(\tau s)$. Substituting the rescaled control
parameters into the flux-velocity Eq.~(\ref{eq:volecity}), we obtain
\begin{equation}
    V_{\tau} = -\mu [\bar{u}_2(s)x+\bar{u}_1(s)] - \tau[\mu U'_{\mathrm{ref}}+D\partial_x \ln\rho_{\tau}].
    \label{eq:volecity_rescale_full}
\end{equation}
When $\tau \to 0$, the contributions from the reference potential and diffusion both vanish, and we retain only the $O(1)$ term in Eq.~(\ref{eq:volecity_rescale_full}),
\begin{equation}
    V_0(x,s)=\alpha(s)x + \beta(s),
\end{equation}
where $\alpha = -\mu \bar{u}_2$ and $\beta = -\mu \bar{u}_1$. Thus the leading-order probability-flux velocity takes an affine form in space.

Using the zeroth-order velocity $V_0$, the characteristic curves of the Fokker--Planck equation satisfy
\begin{equation}
    \frac{\mathrm{d}X_s}{\mathrm{d}s}=\alpha (s) X_s + \beta(s).
    \label{eq:characteristic}
\end{equation}
Each curve starts from a different $x_0$, and solving Eq.~(\ref{eq:characteristic}) gives
\begin{equation}
    X_s(x_0) = A_s x_0 + B_s,
\end{equation}
with $A'_s=\alpha A_s$, $B'_s=\alpha B_s + \beta$, and the initial conditions $A_0=1$, $B_0=0$.
Because $A_s=\exp[\int_0^s \alpha(u)\,\mathrm{d}u]>0$, this affine mapping is order-preserving and pushes the distribution to the right according to $\rho_s(A_s x_0 + B_s)=\rho_0(x_0)/A_s$; the entire distribution is transformed by the mapping, whereas under full control different regions may undergo different displacements. Along the characteristic curves, $\mathrm{d}X_s/\mathrm{d}s=A'_sx_0+B'_s$. Using $\langle X_0\rangle_0=0$ and $\langle X_0^2\rangle_0=m_2$, we rewrite the erasure cost in Eq.~(\ref{eq:irr_work}) as
\begin{equation}
    \begin{aligned}
    \lim_{\tau\to 0} \tau W_{\mathrm{irr}} &= \frac{1}{\mu} \int_0^1 \mathrm{d} s \int_{-\infty}^{+\infty}\mathrm{d}x_0\,\rho_0(x_0)(A_s' x_0 + B_s')^2\\
        &= \frac{1}{\mu}\int_0^1\bigl[m_2(A_s')^2+(B_s')^2\bigr]\,\mathrm{d}s.
    \end{aligned}
    \label{eq:irr_work_limit}
\end{equation}
Here $A_s$ compresses or expands the whole distribution, while $B_s$ shifts it; therefore the optimization of
Eq.~(\ref{eq:irr_work_limit}) transforms into a path optimization in the $(A,B)$ plane, with the metric $\mathrm{d}l^2 = m_2\,\mathrm{d}A^2+\mathrm{d}B^2$. 

We now rephrase the success-rate constraint in the fast-erasure approximation.
For the cumulative distribution function $\mathcal{F}_0(x)=\int_{-\infty}^x \rho_0(y)\,\mathrm{d}y$, we define the target
quantile $q=\mathcal{F}^{-1}_0(1-p)<0$. Because the mapping from the initial to the final state, $X_1 = A_1X_0 + B_1$, is monotonically increasing, Eq.~(\ref{eq:success_rate})
is equivalent to
\begin{equation}
    \begin{aligned}
        A_1 q + B_1 &\geq 0.
    \end{aligned}
    \label{eq:success_condition}
\end{equation}
In the $(\sqrt{m_2}A,B)$ plane, the optimal path from $(\sqrt{m_2},0)$ to $(\sqrt{m_2}A_1,B_1)$ consists of straight lines,
which are generated from the equality condition of the Cauchy--Schwarz inequality; minimizing the irreversible work thus becomes a rigorous convex optimization over the endpoint $(\sqrt{m_2}A_1,B_1)$,
\begin{equation}
    \min_{A_1>0,\,B_1}\frac{1}{\mu}\bigl[m_2(A_1 - 1)^2 + B_1^2\bigr]
    \label{eq:optimal_irr_limit}
\end{equation}
subject to Eq.~(\ref{eq:success_condition}). Because $q<0$, the unconstrained minimum $(A_1,B_1)=(1,0)$ violates Eq.~(\ref{eq:success_condition}), so the true minimum is attained on the constraint boundary $A_1q + B_1 = 0$. Substituting $B_1=-A_1q$ into Eq.~(\ref{eq:optimal_irr_limit}) yields
\begin{equation}
    \Psi(A_1) = m_2 (A_1 - 1)^2 + q^2 A_1^2.
\end{equation}
The condition $\Psi'(A_1)=0$ gives the minimum, so the corresponding optimal terminal coordinates are
\begin{equation}
    A_1^* = \frac{m_2}{m_2 + q^2},\qquad B_1^* = -\frac{m_2q}{m_2 + q^2},
    \label{eq:optimal_endpoint}
\end{equation}
and the optimal paths are
\begin{equation}
    A_s^* = 1 + s(A_1^* - 1),\qquad B_s^* = sB_1^*.
    \label{eq:optimal_paths}
\end{equation}
From Eq.~(\ref{eq:irr_work_limit}), as $\tau\to 0$, $\tau W_{\mathrm{irr}}$ tends to the constant $C_0$,
\begin{equation}
    \begin{aligned}
    C_0 &= \frac{1}{\mu}\bigl[m_2(A_1^* - 1)^2 + (B_1^*)^2\bigr]\\
    &= \frac{m_2q^2}{\mu (m_2 + q^2)}.
    \end{aligned}
    \label{eq:C0}
\end{equation}
We call the optimal protocols in the above fast-erasure limit the zeroth-order protocols, denoted $u_0$.
We note that $C_0$ depends only on $m_2$ and $q$, while the shape of the reference
potential has no influence on it.   

To quantify the extra cost induced by the restricted linear--quadratic trap, we calculate the cost under full control.
When the potential landscape can be freely transformed, one need only transport the required part of the probability mass from the region $(q,0)$ to the right of $0$, forming a peak while leaving the rest unchanged. The full-control cost is then
\begin{equation}
    C_{\mathrm{full}} = \frac{1}{\mu} \int_q^0 x^2\rho_0(x)\,\mathrm{d}x.
\end{equation}
It is clear that $C_{\mathrm{full}} < C_0$; equality is impossible because the linear--quadratic control unavoidably transports extra probability mass. For a typical equilibrium distribution with a finite second moment, as $p\to 1$, $C_0\to m_2/\mu$ and $C_{\mathrm{full}}\to m_2/(2\mu)$, so the ratio $C_0/C_{\mathrm{full}}$ tends to $2$, reflecting the difference between transporting the whole distribution and transporting only half.
\subsection{First-order finite-time correction}
\label{subsec:first-order}
The above results apply only when $\tau$ is close to zero.
Here we derive the first-order correction for erasure at a finite but small duration, using the zeroth-order optimal protocols.
In this regime the free-energy difference is no longer negligible, and the dynamics generally deviates from the zeroth-order affine-transport solution. Expanding Eq.~(\ref{eq:irr_work}) with respect to $\tau$ near $\tau=0$ up to first order gives
\begin{equation}
    \tau \langle W\rangle_{\mathrm{opt}} = C_0 + C_1 \tau + O(\tau^2),
    \label{eq:work_expand}
\end{equation}
where the coefficient $C_0$ has been derived above.
The first-order coefficient $C_1$ comprises two contributions: the first arises from the finite-time deviation from the zeroth-order affine-transport solution, where diffusion and the free-energy difference must also be accounted for; the second stems from the modification of the protocols required to meet the final success-rate constraint.

To calculate $C_1$, we seek an explicit description of the evolution of the part of the distribution that deviates from affine transport. We therefore introduce the coordinate transformation
\begin{equation}
    y = \frac{x-B_s^*}{A_s^*},
\end{equation}
which maps $\rho_{\tau}$ to
\begin{equation}
    \pi_{\tau}(y,s) = A_s^*\rho_{\tau}(A_s^* y + B_s^*,s).
    \label{eq:trans_dis}
\end{equation}
This transformation cancels the affine-transport term in the corresponding part of the Fokker--Planck equation and explicitly retains the influence of the reference potential, which was neglected in the fast-erasure limit. Expanding $\pi_{\tau}$ with respect to $\tau$ near $\tau=0$ gives
\begin{equation}
    \pi_{\tau} = \rho_0 + \tau \pi^{[1]} + O(\tau^2),
    \label{eq:trans_trunacate}
\end{equation}
with the initial condition $\pi^{[1]}(y,0)=0$. Substituting Eq.~(\ref{eq:trans_trunacate}) into the transformed Fokker--Planck equation yields the equation for $\pi^{[1]}$,
\begin{equation}
    \partial_s \pi^{[1]} = -\partial_y\!\bigl[\frac{b_r(A_s^*y+B_s^*)}{A_s^*}\rho_0(y)\bigr] + \frac{D}{(A_s^*)^2}\rho_0''(y),
    \label{eq:trans_tr_eqaution}
\end{equation}
where we have defined $b_r(x)=-\mu U_{\mathrm{ref}}'(x)$. The function $\pi^{[1]}$ is determined solely by the initial distribution, the reference potential, and the zeroth-order optimal paths.

Meanwhile, for the given optimal zeroth-order protocols, the success-rate correction at a finite but small erasure time is
\begin{equation}
    P_0(\tau) = p - r\tau + O(\tau^2),
    \label{eq:success_deviation}
\end{equation}
with
\begin{equation}
    r = -\rho_0(q) \int_0^1 \frac{b_r(A_s^* q + B_s^*)}{A_s^*}\,\mathrm{d}s + D\rho_0'(q)\int_0^1\frac{1}{(A_s^*)^2}\,\mathrm{d}s.
    \label{eq:r}
\end{equation}
If $r>0$, the drift induced by the reference potential and diffusion under the zeroth-order optimal protocols during a finite but small erasure makes the success rate lower than the target value. Conversely, $r<0$ raises the success rate above the target, and the factors influencing $r$ include more than just $m_2$ and $q$.

We can thus quantify the impact of a finite but small erasure duration. The coefficient splits into two parts, $C_1 = \kappa_{\mathrm{dyn}} + \eta_0 r$.
The first part, $\kappa_{\mathrm{dyn}}$, originates from the direct influence of finite time.
Using the definition of work in Eq.~(\ref{eq:irr_work}) and Eq.~(\ref{eq:trans_tr_eqaution}) together with some standard integral identities, we obtain
\begin{equation}
    \kappa_{\mathrm{dyn}} = \kappa_{\mathrm{bdry}} + \kappa_{\mathrm{coup}} + \kappa_{\mathrm{diff}}.
    \label{eq:kappa_dyn_sum}
\end{equation}
The term $\kappa_{\mathrm{bdry}}$ arises from the change in free energy,
\begin{widetext}
\begin{equation}
    \kappa_{\mathrm{bdry}} = \int_{-\infty}^{\infty} \rho_0(y)\bigl[U_{\mathrm{ref}}(A_1^* y+B_1^*)-U_{\mathrm{ref}}(y)\bigr]\,\mathrm{d}y - \frac{D}{\mu}\ln A_1^*,
    \label{eq:kappa_bdry}
\end{equation}
\end{widetext}
the excess dissipation due to the deviation from affine transport is
\begin{widetext}
\begin{equation}
    \kappa_{\mathrm{coup}} = \frac{2A_1^*}{\mu} \int_{-\infty}^{\infty}\rho_0(y)\bigl[(A_1^* - 1)y + B_1^*\bigr]\biggl[\int_0^1\frac{b_r(A_s^* y + B_s^*)}{A_s^*}\,\mathrm{d}s\biggr]\,\mathrm{d}y,
    \label{eq:kappa_coup}
\end{equation}
\end{widetext}
and the diffusive contribution is
\begin{equation}
    \kappa_{\mathrm{diff}} = \frac{2D}{\mu}(A_1^*-1).
    \label{eq:kappa_diff}
\end{equation}
Detailed derivations of the first-order finite-time corrections are provided in Appendix~\ref{app:first-order}.

To account for the finite-time deviation of the success rate under the zeroth-order optimal protocols, we note that $C_0$ depends on the success rate $p$ and define
\begin{equation}
    \eta_0 \equiv \frac{\mathrm{d}C_0}{\mathrm{d}p}=-\frac{2m_2^2 q}{\mu \rho_0(q)(m_2+q^2)^2} > 0.
    \label{eq:eta}
\end{equation}
This quantity captures the sensitivity of the fast-erasure cost to changes in the success rate. Because the zeroth-order optimal protocols produce a finite-time shift of the final success rate by $-r\tau$ relative to the target value $p$, restoring the terminal success-rate constraint requires a corresponding change of $r\tau$ in the success rate through a modification of the protocols. This changes $C_0$ by $(dC_0/dp)r\tau=\eta_0 r\tau$, i.e., an $O(\tau)$ correction.

\section{Examples}
\label{sec:numerical}

\subsection{Comparison with the exact harmonic solution for arbitrary erasure duration}
\label{sec:harmonic-comparison}

For the harmonic reference potential $U_{\mathrm{ref}}(x)=k_0x^2/2$,
the initial equilibrium distribution is
$\rho_0=\mathcal{N}(0,\sigma_0^2)$, where
$\sigma_0^2=k_{\mathrm B}T/k_0$ and $D=\mu k_{\mathrm B}T$.
Defining $z_p=\Phi^{-1}(p)>0$ and $a_p=1+z_p^2$, we obtain
\begin{equation}\label{eq:S44}
    \begin{aligned}
    m_2&=\sigma_0^2,&
    q&=-\sigma_0z_p,\\
    b_r(x)&=-\mu k_0x,&
    \rho_0'(q)&=-\frac{q}{\sigma_0^2}\rho_0(q).
    \end{aligned}
\end{equation}
The zeroth-order optimal affine coefficients become
\begin{equation}\label{eq:S45}
    \begin{aligned}
    A_1^*&=\frac{1}{a_p},&
    B_1^*&=\frac{\sigma_0z_p}{a_p},\\
    A_s^*&=1-\frac{z_p^2}{a_p}s,&
    B_s^*&=\frac{\sigma_0z_p}{a_p}s.
    \end{aligned}
\end{equation}
The leading coefficient in Eq.~(\ref{eq:C0}) therefore reduces to
\begin{equation}\label{eq:S46}
    C_0=\frac{m_2q^2}{\mu(m_2+q^2)}
    =\frac{\sigma_0^2z_p^2}{\mu a_p}.
\end{equation}

To evaluate $C_1$, we first compute
\begin{equation}\label{eq:S47}
    \begin{aligned}
    I_B\equiv\int_0^1\frac{B_s^*}{A_s^*}\,ds
    &=\frac{\sigma_0(a_p\ln a_p-z_p^2)}{z_p^3},\\
    I_2\equiv\int_0^1\frac{ds}{(A_s^*)^2}
    &=a_p,
    \end{aligned}
\end{equation}
so that
\begin{equation}\label{eq:S48}
    \int_0^1\frac{b_r(A_s^*y+B_s^*)}{A_s^*}\,ds
    =-\mu k_0(y+I_B).
\end{equation}
Using $\langle y\rangle_0=0$,
$\langle y^2\rangle_0=\sigma_0^2$, and $k_0\sigma_0^2=k_{\mathrm B}T$,
Eqs.~(\ref{eq:kappa_bdry}), (\ref{eq:kappa_coup}), and (\ref{eq:kappa_diff}) give, respectively,
\begin{equation}\label{eq:S49}
    \begin{aligned}
    \kappa_{\mathrm{bdry}}
    &=\frac{k_0}{2}
    \left[\sigma_0^2((A_1^*)^2-1)+(B_1^*)^2\right]
    -\frac{D}{\mu}\ln A_1^*\\
    &=k_{\mathrm B}T\left(\ln a_p-\frac{z_p^2}{2a_p}\right),
    \end{aligned}
\end{equation}
\begin{equation}\label{eq:S50}
    \begin{aligned}
    \kappa_{\mathrm{coup}}
    &=-2k_0A_1^*
    \left[(A_1^*-1)\sigma_0^2+B_1^*I_B\right]\\
    &=\frac{2k_{\mathrm B}T}{a_p}
    \left(1-\frac{\ln a_p}{z_p^2}\right),
    \end{aligned}
\end{equation}
and
\begin{equation}\label{eq:S51}
    \kappa_{\mathrm{diff}}
    =\frac{2D}{\mu}(A_1^*-1)
    =-\frac{2k_{\mathrm B}Tz_p^2}{a_p}.
\end{equation}
Similarly, Eqs.~(\ref{eq:r}) and (\ref{eq:eta}) yield
\begin{equation}\label{eq:S52}
    \begin{aligned}
    r
    &=\mu k_0\rho_0(q)(q+I_B)+D\rho_0'(q)a_p\\
    &=\frac{\mu k_0\sigma_0a_p\rho_0(q)}{z_p^3}
    \left(\ln a_p-z_p^2+z_p^4\right),
    \end{aligned}
\end{equation}
\begin{equation}\label{eq:S53}
    \eta_0
    =-\frac{2m_2^2q}{\mu\rho_0(q)(m_2+q^2)^2}
    =\frac{2\sigma_0z_p}{\mu\rho_0(q)a_p^2}.
\end{equation}
Hence,
\begin{equation}\label{eq:S54}
    \eta_0r
    =\frac{2k_{\mathrm B}T}{a_p}
    \left(\frac{\ln a_p}{z_p^2}-1+z_p^2\right).
\end{equation}
Substituting these terms into $C_1 = \kappa_{\mathrm{dyn}} + \eta_0 r$, we obtain
\begin{equation}\label{eq:S55}
    \begin{aligned}
    C_1
    &=\kappa_{\mathrm{bdry}}+\kappa_{\mathrm{coup}}
    +\kappa_{\mathrm{diff}}+\eta_0r\\
    &=k_{\mathrm B}T\left(\ln a_p-\frac{z_p^2}{2a_p}\right).
    \end{aligned}
\end{equation}

With $\beta=(k_{\mathrm B}T)^{-1}$ and $\theta=\mu k_0\tau$,
the general fast-erasure expansion becomes
\begin{equation}\label{eq:S56}
    \beta\langle W\rangle_{\mathrm{opt}}
    =\frac{z_p^2}{a_p\theta}
    +\left(\ln a_p-\frac{z_p^2}{2a_p}\right)
    +O(\theta).
\end{equation}
This expression reproduces the leading and constant terms in the short-time
expansion of the exact harmonic solution, Eq.~(35) of our companion paper \cite{jiang2026harmonic}, confirming agreement of $C_0$ and $C_1$.

\subsection{Numerical methods and short-time theoretical tests}
\label{sec:114514}
In the usual finite-time regime, nonlinearity and the breakdown of the fast-erasure approximation render theoretical optimization intractable. In this section we therefore use numerical methods to obtain optimal protocols and to verify our fast-erasure approximation and first-order correction.

To verify our analytical results, we adopt the classical quartic double-well potential
\begin{equation}
    U(x;g,f) = 4E_b\biggl(\frac{x^4}{4}-\frac{gx^2}{2}-fx\biggr),
    \label{eq:quartic_potential}
\end{equation}
where $g$ and $f$ are dimensionless control parameters. When $g=1$ and $f=0$, the potential reduces to the reference potential $U_{\mathrm{ref}}(x)=E_b(x^4-2x^2)$, which has two minima at $x=\pm 1$ and a barrier of height $E_b$. The parameter $g(t)$ controls the barrier height and $f(t)$ controls the tilt, corresponding to $u_1=-4E_b f$ and $u_2=-4E_b(g-1)$. Defining the rescaled protocols $G(s)=\tau g(\tau s)$ and $F(s)=\tau f(\tau s)$ and inserting these into the Fokker--Planck equation gives
\begin{equation}
    \partial_s \rho = -\partial_x\bigl\{4\mu E_b [G(s)x+F(s)-\tau x^3]\rho\bigr\} + D\tau \partial_x^2\rho.
    \label{eq:quartic_fp}
\end{equation}
For our simulations we set $E_b=4$, $\mu=D=0.1$, $k_{\mathrm{B}}T=1$, and $p=0.999$.

In our numerical optimization we use cubic B-splines to parameterize the protocols.
The optimization proceeds in two steps. First, we use threshold-accepting Monte Carlo~\cite{Dueck1990} to coarsely search for candidate protocols with low work that satisfy the success-rate constraint. In the next step, we refine the protocols by constrained gradient descent~\cite{Engel2023,Zhong2022,AbsilMalick2012,Rosen1961}, evaluating the work and its gradient from Langevin trajectory simulations and the success rate and its gradient from the Fokker--Planck equation. Further algorithmic details and numerical verification procedures are given in Appendix~\ref{app:numerical}.

To test the accuracy of our first-order correction, we define the following three quantities:
\begin{equation}
    r^{\mathrm{num}} = \frac{p-P_0(\tau)}{\tau},
    \label{eq:r_num}
\end{equation}
\begin{equation}
    \kappa_{\mathrm{dyn}}^{\mathrm{num}}(\tau) = \frac{\tau \langle W\rangle_0 - C_0}{\tau},
    \label{eq:kappa_dyn_num}
\end{equation}
\begin{equation}
    \kappa_{\mathrm{suc}}^{\mathrm{num}} = \frac{\tau \langle W\rangle-\tau \langle W\rangle_0}{\tau}.
    \label{eq:kappa_suc_num}
\end{equation}
Here $\langle W\rangle_0$ denotes the work computed under the fixed zeroth-order optimal protocols, while $\langle W\rangle$ in Eq.~(\ref{eq:kappa_suc_num}) denotes the work after adjusting those protocols to restore the target success rate; see Appendix~\ref{app:numerical-verification}. The quantity $r^{\mathrm{num}}$ characterizes the effect of the finite but small erasure time on the final success rate; $\kappa_{\mathrm{dyn}}^{\mathrm{num}}$ quantifies the finite-time work correction under the fixed zeroth-order optimal protocols; and $\kappa_{\mathrm{suc}}^{\mathrm{num}}$ represents the work cost required to restore the success rate by tuning the protocols.

\section{Results and discussion}
\label{sec:results}

\begin{figure}[htbp]
  \centering
  \includegraphics[width=0.95\linewidth]{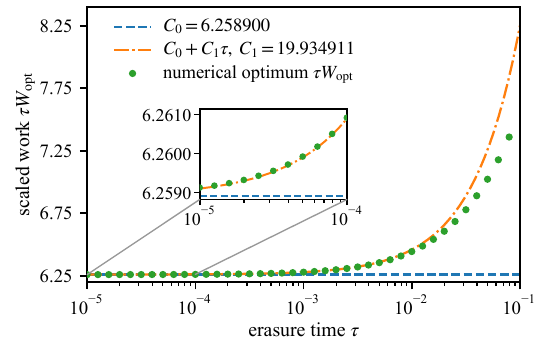} \captionsetup{justification=raggedright,singlelinecheck=false}
  \caption{Comparison of scaled optimal work with the fast-erasure approximation. Green circles denote numerical results, the blue dashed line is the constant $C_0$, and the orange dash-dotted line is the first-order correction $C_0 + C_1\tau$; inset: zoom of the fast regime from $\tau=10^{-5}$ to $\tau=10^{-4}$.}
  \label{fig:work_compare}
\end{figure}

We now present the fast-erasure approximation and compare it with numerical results for the quartic double-well potential, including the zeroth-order prediction and the first-order correction. For comparison, we plot $\tau W_{\mathrm{min}}$, $C_0$, and $C_0 + C_1\tau$ in Fig.~\ref{fig:work_compare}.
% ========== 图 1：最优功与快速擦除近似对比 ==========
As shown in Fig.~\ref{fig:work_compare}, both the zeroth-order and first-order predictions are accurate in the fast-erasure regime, while an observable deviation of the zeroth-order prediction first appears as $\tau$ increases; eventually both asymptotic predictions fail, with the zeroth-order result underestimating the optimal work while the first-order approximation overestimates it. 
  
\begin{figure}[htbp]
  \centering
  \includegraphics[width=1\linewidth]{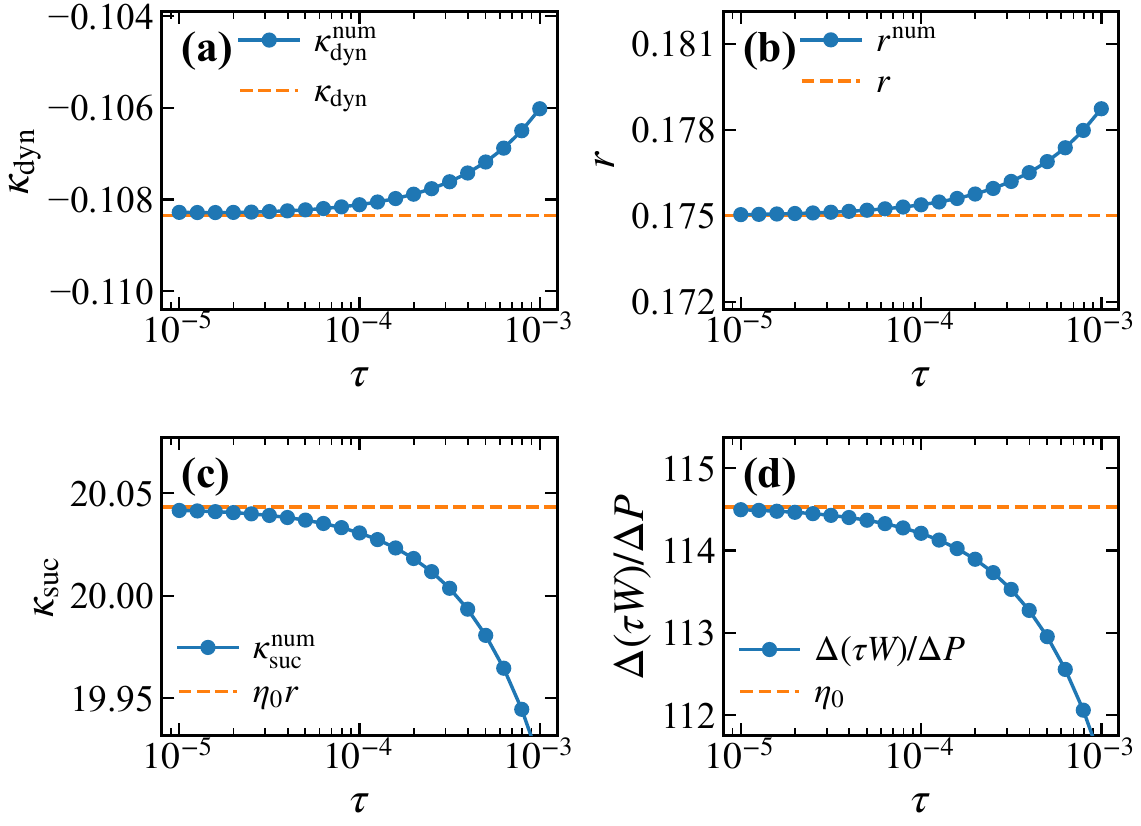} \captionsetup{justification=raggedright,singlelinecheck=false}
  \caption{Coefficients in the first-order correction. Blue circles with solid lines denote numerical results, while orange dashed lines denote theoretical results.
(a)~Work-correction coefficient at fixed zeroth-order optimal protocols for short but finite durations, $\kappa_{\mathrm{dyn}}^{\mathrm{num}}$, compared with $\kappa_{\mathrm{dyn}}$;
(b)~first-order coefficient of the success-rate deviation, $r^{\mathrm{num}}$, compared with $r$;
(c)~work-correction coefficient for restoring the target success rate, $\kappa_{\mathrm{suc}}^{\mathrm{num}}$, compared with $\eta_0 r$;
(d)~ratio $\Delta(\tau W)/\Delta P$, compared with $\eta_0$.
In the limit $\tau\to 0$, the numerical quantities converge to their corresponding theoretical values.}
  \label{fig:coefficient_compare}
\end{figure}

We then examine the first-order correction coefficients $r$, $\kappa_{\mathrm{dyn}}$, $\eta_0 r$, and $\eta_0$, and plot the corresponding four quantities for comparison with numerical results. As shown in Fig.~\ref{fig:coefficient_compare}, all quantities agree well provided the erasure duration is sufficiently short.
% ========== 图 2：一阶修正系数 ==========

\begin{figure}[htbp]
  \centering
  % 若不想重命名，可使用下面一行（需加 \usepackage{grffile}）：
  \includegraphics[width=0.95\linewidth]{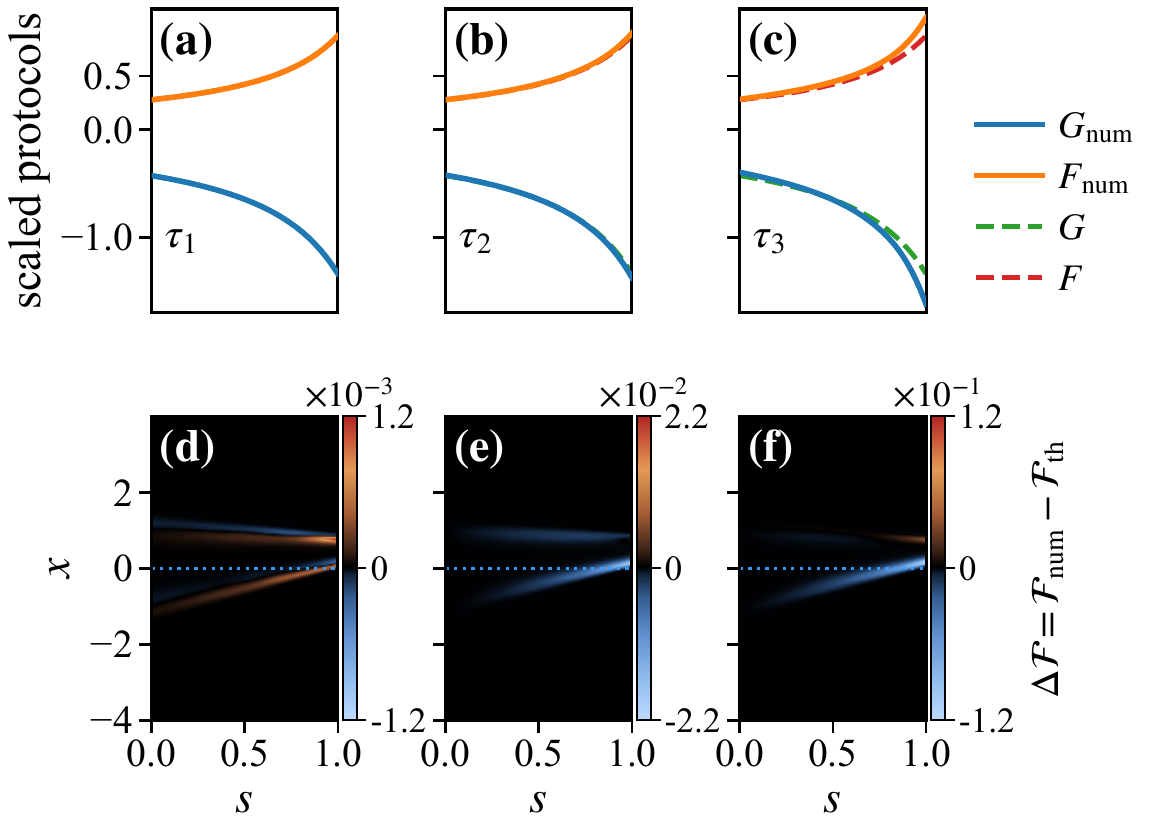} \captionsetup{justification=raggedright,singlelinecheck=false}
  \caption{Comparison of optimal protocols at short but finite durations with the affine-transport approximation. The three columns correspond to $\tau_1=10^{-5}$, $\tau_2=3.98\times10^{-3}$, and $\tau_3=3.16\times10^{-2}$, respectively. The first row shows the scaled optimal protocols $G_{\mathrm{num}}$ and $F_{\mathrm{num}}$ and the corresponding zeroth-order approximate protocols $G$ and $F$; the second row shows the difference of the cumulative distributions, $\Delta {\mathcal{F}}=\mathcal{F}_{\mathrm{num}}-\mathcal{F}_{\mathrm{th}}$, with dotted lines indicating the logical-state boundaries.}
  \label{fig:distribution_compare}
\end{figure}

To examine these finite-time deviations, Fig.~\ref{fig:distribution_compare} compares the numerical optimal protocols and probability distributions with the zeroth-order affine-transport prediction at three erasure durations. In the first row, the numerical optimal protocols and the zeroth-order approximate ones are almost indistinguishable in the fast-erasure regime [Fig.~\ref{fig:distribution_compare}(a)], whereas they deviate for larger $\tau$ [Fig.~\ref{fig:distribution_compare}(c)]. We also plot the difference between the cumulative distributions of the numerical optimal solution and the zeroth-order prediction, $\Delta\mathcal{F}=\mathcal{F}_{\mathrm{num}}-\mathcal{F}_{\mathrm{th}}$. The larger deviation observed in Fig.~\ref{fig:distribution_compare}(f) shows that the optimal distribution increasingly departs from the zeroth-order affine-transport prediction as the erasure duration increases.
% ========== 图 3：最优协议与分布偏差 ==========
% 请将图片文件 "PRE_Fig3 (1).pdf" 重命名为 "Fig3.pdf"，否则需取消下行注释

\section{Conclusion}
\label{sec:conclusion}
% Conclusion: to be filled paragraph by paragraph.

In this paper, we investigated restricted-control erasure in an overdamped system with time-dependent linear and quadratic additions to a reference potential and theoretically derived the optimal protocols. In the fast-erasure limit we found that the optimal control causes the leading-order probability-flux velocity to take an affine form, so that the whole distribution shifts and scales. The corresponding optimal irreversible work depends on time as $\tau^{-1}$, with a coefficient $C_0$ that depends only on the second moment $m_2$ of the initial distribution and the quantile value at the target success rate. Comparison with optimal full control shows that the extra dissipation originates from unnecessary transport of probability mass that does not need to cross the logical-state boundary.

We also provide a first-order correction to the scaled optimal mean work $\tau\langle W\rangle$ for short but finite erasure durations. The correction coefficient $C_1$ comprises two parts: $\kappa_{\mathrm{dyn}}$, which includes the change in nonequilibrium free energy and the corrections to irreversible work due to the reference potential and diffusion under the zeroth-order optimal protocols; and $\eta_0 r$, which stems from restoring the success rate by adjusting the protocols. We verified our predictions on a classical quartic double-well potential using numerical methods and observed the expected agreement at short durations, with deviations emerging at longer durations. The optimal distribution also increasingly departs from the zeroth-order affine-transport prediction as the erasure duration increases, and redistribution within the target logical state is observed.

These results clarify how control restrictions shape finite-time thermodynamic limits through the probability transport they allow. The higher cost relative to full control shows that the shared inverse-duration scaling does not by itself determine the minimum dissipation: its prefactor also reflects the restrictions on transport. At fixed mobility, the dependence of the leading cost on only two initial-state statistics reveals a universal structure in the fast-erasure limit, while the finite-time correction shows how details of the reference potential enter beyond this limit. Extending this analysis to other classes of control potentials could further clarify how accessible probability transport determines minimum dissipation in nonequilibrium processes.

\begin{acknowledgments}
This work was financially
supported by the National Natural Science Foundation
of China (Grants No. 12475039), Guangdong Basic and Applied Basic
Research Foundation (Grant No. 2025A1515010350) and the Fundamental Research Funds for the Central Universities (Grant No. 20720262015).
\end{acknowledgments}

% ======================================================================

\appendix

\section{Derivation of the first-order finite-time correction}
\label{app:first-order}

This Appendix gives the detailed calculation underlying
Sec.~\ref{subsec:first-order}. We use the model and the optimal affine
paths defined in Sec.~\ref{sec:theory}, without repeating the
fast-erasure minimization of Sec.~\ref{subsec:fast-erasure}.

\subsection{First-order expansion in comoving coordinates}
\label{app:comoving}

For the fixed zeroth-order optimal protocol, the total potential is
\begin{equation}
 U_{\tau}^{[0]}(x,s)=U_{\mathrm{ref}}(x)
 +\frac{1}{\tau}U_c^{[0]}(x,s),
 \label{eq:app_fixed_potential}
\end{equation}
where $U_c^{[0]}$ is the linear--quadratic control potential that
generates the optimal paths in Eq.~(\ref{eq:optimal_paths}). Define
\begin{equation}
 \begin{aligned}
 b_0(x,s)&=-\mu\partial_x U_c^{[0]}(x,s),\\
 b_r(x)&=-\mu U_{\mathrm{ref}}'(x).
 \end{aligned}
 \label{eq:app_drifts}
\end{equation}
In the scaled time $s=t/\tau$, the Fokker--Planck equation becomes
\begin{equation}
 \partial_s\rho_{\tau}
 =-\partial_x[(b_0+\tau b_r)\rho_{\tau}]
 +D\tau\partial_x^2\rho_{\tau}.
 \label{eq:app_fp}
\end{equation}

Introduce the comoving coordinate and density
\begin{equation}
 x=A_s^*y+B_s^*,\qquad
 \pi_{\tau}(y,s)=A_s^*\rho_{\tau}(A_s^*y+B_s^*,s).
 \label{eq:app_comoving_transform}
\end{equation}
Since the zeroth-order affine characteristics satisfy
\begin{equation}
 (A_s^*)'y+(B_s^*)'
 =b_0(A_s^*y+B_s^*,s),
 \label{eq:app_affine_identity}
\end{equation}
all $O(1)$ advection terms cancel after the coordinate transformation.
The remaining evolution is
\begin{equation}
 \begin{aligned}
 \partial_s\pi_{\tau}
 ={}&-\tau\partial_y\!\left[
 \frac{b_r(A_s^*y+B_s^*)}{A_s^*}\pi_{\tau}\right]\\
 &+\frac{D\tau}{(A_s^*)^2}\partial_y^2\pi_{\tau}.
 \end{aligned}
 \label{eq:app_comoving_fp}
\end{equation}
Expanding
\begin{equation}
 \pi_{\tau}(y,s)=\rho_0(y)+\tau\pi^{[1]}(y,s)+O(\tau^2)
 \label{eq:app_pi_expansion}
\end{equation}
and using $\pi^{[1]}(y,0)=0$ gives
\begin{equation}
 \partial_s\pi^{[1]}
 =-\partial_y\!\left[
 \frac{b_r(A_s^*y+B_s^*)}{A_s^*}\rho_0(y)\right]
 +\frac{D}{(A_s^*)^2}\rho_0''(y),
 \label{eq:app_pi1_equation}
\end{equation}
which is Eq.~(\ref{eq:trans_tr_eqaution}) in the main text.

For later use, define
\begin{equation}
 \mathcal{R}_s(y)
 \equiv\int_0^s
 \frac{b_r(A_u^*y+B_u^*)}{A_u^*}\,\mathrm{d}u,
 \qquad
 \mathcal{I}_s
 \equiv\int_0^s\frac{\mathrm{d}u}{(A_u^*)^2}.
 \label{eq:app_RI}
\end{equation}
Direct integration of Eq.~(\ref{eq:app_pi1_equation}) then yields
\begin{equation}
 \pi^{[1]}(y,s)
 =-\partial_y[\rho_0(y)\mathcal{R}_s(y)]
 +D\rho_0''(y)\mathcal{I}_s.
 \label{eq:app_density_correction}
\end{equation}
Thus the first departure from affine transport is generated by two
sources that are absent in the strict fast limit: the reference-potential
drift and diffusion.

\subsection{Success-rate deviation under the fixed zeroth-order protocol}
\label{app:success-deviation}

At the optimal endpoint, $A_1^*q+B_1^*=0$ and $A_1^*>0$.
Therefore $x>0$ at $s=1$ is equivalent to $y>q$, and
\begin{equation}
 P_0(\tau)
 =\int_q^\infty\pi_{\tau}(y,1)\,\mathrm{d}y
 =p+\tau\int_q^\infty\pi^{[1]}(y,1)\,\mathrm{d}y
 +O(\tau^2).
 \label{eq:app_success_integral}
\end{equation}
Using Eq.~(\ref{eq:app_density_correction}) and taking the boundary
terms at $y\to\infty$ to vanish gives
\begin{equation}
 \int_q^\infty\pi^{[1]}(y,1)\,\mathrm{d}y
 =\rho_0(q)\mathcal{R}_1(q)
 -D\rho_0'(q)\mathcal{I}_1.
 \label{eq:app_success_pi1}
\end{equation}
Hence
\begin{equation}
 P_0(\tau)=p-\tau r+O(\tau^2),
 \label{eq:app_success_expansion}
\end{equation}
with
\begin{equation}
 r=-\rho_0(q)\int_0^1
 \frac{b_r(A_s^*q+B_s^*)}{A_s^*}\,\mathrm{d}s
 +D\rho_0'(q)\int_0^1\frac{\mathrm{d}s}{(A_s^*)^2},
 \label{eq:app_r_explicit}
\end{equation}
identical to Eq.~(\ref{eq:r}).

The first term samples the reference drift along the image of the
threshold point,
\begin{equation}
 x_s=A_s^*q+B_s^*,
\end{equation}
while the second is the diffusive flux across that moving threshold.
Consequently, unlike $C_0$, which depends only on $m_2$ and $q$, the
coefficient $r$ probes local information about $\rho_0$ and the spatial
structure of $U_{\mathrm{ref}}$ along the zeroth-order transport path.

\subsection{Work correction under the fixed zeroth-order protocol}
\label{app:work-correction}

Let $\langle W\rangle_0$ denote the work under the fixed zeroth-order
optimal protocol. From Eq.~(\ref{eq:irr_work}),
\begin{equation}
 \tau\langle W\rangle_0
 =\tau W_{\mathrm{irr},0}
 +\tau\Delta\mathscr{F}_{\mathrm{neq}}
 =C_0+\tau\kappa_{\mathrm{dyn}}+O(\tau^2).
 \label{eq:app_fixed_work}
\end{equation}
We evaluate the free-energy and irreversible-work pieces separately.

The leading terminal density is the affine image
\begin{equation}
 \rho_f^{[0]}(x)=\frac{1}{A_1^*}
 \rho_0\!\left(\frac{x-B_1^*}{A_1^*}\right).
 \label{eq:app_final_density}
\end{equation}
Because the potential is $U_{\mathrm{ref}}$ both before and after the
operation, changing variables $x=A_1^*y+B_1^*$ gives
\begin{widetext}
\begin{equation}
 \begin{aligned}
 \Delta\mathscr{F}_{\mathrm{neq}}^{[0]}
 ={}&
 \int_{-\infty}^{\infty}\rho_0(y)
 \left[U_{\mathrm{ref}}(A_1^*y+B_1^*)
 -U_{\mathrm{ref}}(y)\right]\,\mathrm{d}y\\
 &+k_{\mathrm B}T
 \int_{-\infty}^{\infty}\rho_0(y)
 \ln\!\left(\frac{1}{A_1^*}\right)\,\mathrm{d}y\\
 ={}&
 \int_{-\infty}^{\infty}\rho_0(y)
 \left[U_{\mathrm{ref}}(A_1^*y+B_1^*)
 -U_{\mathrm{ref}}(y)\right]\,\mathrm{d}y
 -\frac{D}{\mu}\ln A_1^*.
 \end{aligned}
 \label{eq:app_boundary_derivation}
\end{equation}
\end{widetext}
This is precisely $\kappa_{\mathrm{bdry}}$ in
Eq.~(\ref{eq:kappa_bdry}).

For the irreversible part, write the rescaled probability-flux velocity
in comoving coordinates as
\begin{equation}
 V_{\tau}(A_s^*y+B_s^*,s)
 =V_0(y)+\tau V_1(y,s)+O(\tau^2).
 \label{eq:app_velocity_expansion}
\end{equation}
Since $A_s^*$ and $B_s^*$ are linear in $s$,
\begin{equation}
 V_0(y)=ay+b,\qquad
 a\equiv A_1^*-1,\qquad b\equiv B_1^*,
 \label{eq:app_velocity_zero}
\end{equation}
and $A_s^*=1+as$, $B_s^*=bs$. Equation~(\ref{eq:volecity_rescale_full})
gives
\begin{equation}
 V_1(y,s)
 =b_r(A_s^*y+B_s^*)
 -\frac{D}{A_s^*}\partial_y\ln\rho_0(y).
 \label{eq:app_velocity_one}
\end{equation}
Here the $O(\tau)$ correction to $\ln\pi_\tau$ is not needed in
$V_1$, because the diffusive term in
Eq.~(\ref{eq:volecity_rescale_full}) already carries an explicit
factor of $\tau$.

Using $\mathrm{d}x=A_s^*\mathrm{d}y$ and
$\rho_\tau\,\mathrm{d}x=\pi_\tau\,\mathrm{d}y$, the irreversible work
becomes
\begin{widetext}
\begin{equation}
 \begin{aligned}
 \tau W_{\mathrm{irr},0}
 ={}&\frac{1}{\mu}\int_0^1\!\mathrm{d}s
 \int_{-\infty}^{\infty}\!\mathrm{d}y\,
 \pi_\tau
 \left(V_0+\tau V_1\right)^2+O(\tau^2)\\
 ={}&C_0+\frac{\tau}{\mu}\int_0^1\!\mathrm{d}s
 \int_{-\infty}^{\infty}\!\mathrm{d}y\,
 \left[\pi^{[1]}V_0^2+2\rho_0V_0V_1\right]
 +O(\tau^2).
 \end{aligned}
 \label{eq:app_irr_first_order}
\end{equation}
\end{widetext}
The two terms inside the square brackets must be kept together:
$\pi^{[1]}V_0^2$ accounts for the deformation of the density, whereas
$2\rho_0V_0V_1$ accounts for the correction to the flux velocity.

We first collect all terms proportional to the reference drift.
Using Eq.~(\ref{eq:app_density_correction}) and integrating once by
parts in $y$,
\begin{widetext}
\begin{equation}
 \begin{aligned}
 \kappa_r
 =\frac{2}{\mu}\int_{-\infty}^{\infty}\!\mathrm{d}y\,
 \rho_0(y)V_0(y)
 \biggl[
 a\int_0^1\mathcal{R}_s(y)\,\mathrm{d}s
 +\int_0^1 b_r(A_s^*y+B_s^*)\,\mathrm{d}s
 \biggr].
 \end{aligned}
 \label{eq:app_kappa_r_intermediate}
\end{equation}
\end{widetext}
Fubini's theorem gives
\begin{equation}
 \int_0^1\mathcal{R}_s(y)\,\mathrm{d}s
 =\int_0^1(1-s)
 \frac{b_r(A_s^*y+B_s^*)}{A_s^*}\,\mathrm{d}s.
 \label{eq:app_R_integral}
\end{equation}
Since
\begin{equation}
 1+\frac{a(1-s)}{A_s^*}
 =\frac{A_1^*}{A_s^*},
 \label{eq:app_affine_identity_coup}
\end{equation}
Eq.~(\ref{eq:app_kappa_r_intermediate}) reduces to
\begin{widetext}
\begin{equation}
 \kappa_r
 =\frac{2A_1^*}{\mu}
 \int_{-\infty}^{\infty}\rho_0(y)
 \bigl[(A_1^*-1)y+B_1^*\bigr]
 \left[\int_0^1
 \frac{b_r(A_s^*y+B_s^*)}{A_s^*}\,\mathrm{d}s\right]
 \mathrm{d}y
 =\kappa_{\mathrm{coup}}.
 \label{eq:app_kappa_coup_derivation}
\end{equation}
\end{widetext}
Thus Eq.~(\ref{eq:kappa_coup}) contains both the density deformation
generated by the reference drift and the direct reference-drift
correction to the flux velocity.

The remaining terms are diffusive. For the equilibrium densities
considered here, the boundary terms vanish, and two integrations by
parts give
\begin{equation}
 \int_{-\infty}^{\infty}\rho_0''(y)V_0^2(y)\,\mathrm{d}y
 =2a^2,
 \qquad
 \int_{-\infty}^{\infty}V_0(y)\rho_0'(y)\,\mathrm{d}y
 =-a.
 \label{eq:app_diff_integrals}
\end{equation}
Therefore
\begin{equation}
 \kappa_D
 =\frac{2D}{\mu}
 \left[
 a^2\int_0^1\mathcal{I}_s\,\mathrm{d}s
 +a\int_0^1\frac{\mathrm{d}s}{A_s^*}
 \right].
 \label{eq:app_kappa_D_intermediate}
\end{equation}
Again using Fubini's theorem,
\begin{equation}
 \int_0^1\mathcal{I}_s\,\mathrm{d}s
 =\int_0^1\frac{1-s}{(A_s^*)^2}\,\mathrm{d}s.
 \label{eq:app_I_integral}
\end{equation}
The linear form $A_s^*=1+as$ then implies
\begin{equation}
 \begin{aligned}
 &a^2\int_0^1\frac{1-s}{(A_s^*)^2}\,\mathrm{d}s
 +a\int_0^1\frac{\mathrm{d}s}{A_s^*}\\
 &\qquad
 =a\int_0^1
 \left[\frac{1}{A_s^*}
 +\frac{a(1-s)}{(A_s^*)^2}\right]\mathrm{d}s
 =aA_1^*\int_0^1\frac{\mathrm{d}s}{(A_s^*)^2}
 =a.
 \end{aligned}
 \label{eq:app_diff_identity}
\end{equation}
Hence
\begin{equation}
 \kappa_D=\frac{2D}{\mu}(A_1^*-1)
 =\kappa_{\mathrm{diff}},
 \label{eq:app_kappa_diff_derivation}
\end{equation}
which proves Eq.~(\ref{eq:kappa_diff}). Combining the boundary,
reference-drift, and diffusive pieces yields
\begin{equation}
 \kappa_{\mathrm{dyn}}
 =\kappa_{\mathrm{bdry}}
 +\kappa_{\mathrm{coup}}
 +\kappa_{\mathrm{diff}},
\end{equation}
as stated in Eq.~(\ref{eq:kappa_dyn_sum}).

\subsection{Restoring the success-rate constraint}
\label{app:success-restoration}

The fixed zeroth-order optimal protocol is designed to have the
leading success rate $p$, but Eq.~(\ref{eq:app_success_expansion})
shows that at finite $\tau$ it actually produces
$p-r\tau+O(\tau^2)$. To compensate this deviation at first order, the
leading affine problem must therefore be solved for the slightly shifted
target
\begin{equation}
 p_\tau=p+r\tau+O(\tau^2).
 \label{eq:app_shifted_target}
\end{equation}
Indeed, the dynamical correction evaluated at this shifted protocol
differs from $r$ only by $O(\tau)$, so its contribution to the success
probability differs only at $O(\tau^2)$.

Let $C_0(p)$ denote the minimum leading cost as a function of the target
success rate. Using $\eta_0=\mathrm{d}C_0/\mathrm{d}p$ from
Eq.~(\ref{eq:eta}),
\begin{equation}
 C_0(p_\tau)
 =C_0(p)+\eta_0r\,\tau+O(\tau^2).
 \label{eq:app_C0_shift}
\end{equation}
To see why further optimization does not change this first-order
cost, let $\mathcal{J}$ denote the leading action in
Eq.~(\ref{eq:irr_work_limit}) and $P^{[0]}$ the leading terminal
success probability. At the constrained zeroth-order optimum,
stationarity gives
\begin{equation}
 \delta\mathcal{J}=\eta_0\,\delta P^{[0]}.
 \label{eq:app_constrained_stationarity}
\end{equation}
Thus restoring $\delta P^{[0]}=r\tau$ costs $\eta_0r\tau$ to
first order, while variations tangent to the success-rate constraint
have vanishing first variation of the leading action.

At the same time, the $O(\tau)$ change of the protocol modifies
$\kappa_{\mathrm{dyn}}$ only by $O(\tau)$; because
$\kappa_{\mathrm{dyn}}$ itself is multiplied by $\tau$ in the scaled
work, this affects $\tau\langle W\rangle$ only at $O(\tau^2)$.
Therefore $\kappa_{\mathrm{dyn}}$ may be evaluated on the fixed
zeroth-order optimum when determining the first-order coefficient.
Combining Eqs.~(\ref{eq:app_fixed_work}) and
(\ref{eq:app_C0_shift}) gives
\begin{equation}
 \tau\langle W\rangle_{\mathrm{opt}}
 =C_0+\tau\left(\kappa_{\mathrm{dyn}}+\eta_0r\right)
 +O(\tau^2),
 \label{eq:app_optimal_work}
\end{equation}
and hence
\begin{equation}
 C_1=\kappa_{\mathrm{dyn}}+\eta_0r.
\end{equation}
This separates the direct finite-time dynamical correction from the
additional cost of restoring the terminal success-rate constraint.

\section{Numerical implementation and verification}
\label{app:numerical}

\subsection{Protocol representation and constrained optimization}
\label{app:optimization}

We use the quartic double-well potential in
Eq.~(\ref{eq:quartic_potential}) and the parameters specified in
Sec.~\ref{sec:114514}: $E_b=4$, $\mu=D=0.1$,
$k_{\mathrm{B}}T=1$, and $p=0.999$. With
$G(s)=\tau g(\tau s)$ and $F(s)=\tau f(\tau s)$, the rescaled
Langevin equation is
\begin{equation}
 \mathrm{d}X_s
 =4\mu E_b[G(s)X_s+F(s)-\tau X_s^3]\,\mathrm{d}s
 +\sqrt{2D\tau}\,\mathrm{d}\mathcal{B}_s.
 \label{eq:app_quartic_langevin}
\end{equation}
The protocols $G$ and $F$ are represented by cubic B-splines,
with endpoint jumps allowed at the start and end of erasure.

For each $\tau$, a threshold-accepting Monte Carlo search first
finds low-work candidate protocols satisfying the success-rate
constraint~\cite{Dueck1990}. During the subsequent refinement,
the work and its gradient are evaluated from Langevin trajectory
simulations using common random numbers. The success rate and
its gradient are obtained from the deterministic Fokker--Planck
equation, Eq.~(\ref{eq:quartic_fp}). At each step, the work-descent
direction is projected onto the local tangent space of the
success-rate constraint. After a finite trial step, the protocol
is corrected along the constraint-normal direction to restore
$P_{\mathrm{suc}}(\tau)=p$ before acceptance. Related gradient-based
protocol optimization and constrained projection methods are
discussed in Refs.~\cite{Engel2023,Zhong2022,AbsilMalick2012,Rosen1961}.

\subsection{Fokker--Planck evaluation and numerical resolution}
\label{app:fp-evaluation}

Equation~(\ref{eq:quartic_fp}) is solved on a spatial domain
truncated at $x=\pm4$, using fourth-order central differences for
spatial derivatives and fourth-order Runge--Kutta time integration.
The density is negligible near the outer boundaries. In the
success-probability evaluator used for protocol optimization, the
two outermost nodes at each end are fixed to zero; the deterministic
evaluator for the first-order tests uses a periodic continuation of
the finite-difference stencil. The initial density is the equilibrium
distribution of the reference potential specified in
Sec.~\ref{sec:114514}.

The terminal success probability is evaluated by trapezoidal
quadrature over $x>0$, including the logical boundary $x=0$.
For the deterministic work evaluation in the first-order tests,
the first and second moments are averaged over the endpoints of
each time step and multiplied by the corresponding protocol
increments. The work of the initial and final potential jumps is
included separately. These observable evaluations have second-order
quadrature errors, despite the fourth-order density evolution.

To reduce discretization error, the main optimization protocols
are reevaluated on two space--time meshes,
$4096\times16384$ and $8192\times32768$. The success probability
is combined using second-order Richardson extrapolation,
\begin{equation}
 R_{\mathrm{ext}}
 =\frac{4R_{\mathrm{fine}}-R_{\mathrm{coarse}}}{3}.
 \label{eq:app_richardson}
\end{equation}
The verification of the first-order correction uses the finer meshes
$16384\times65536$ and $32768\times131072$, with the same
extrapolation applied to both the success probability and the scaled
mean work. In each pair, both the spatial and temporal resolutions
are doubled.

For the parameters above, the initial equilibrium distribution
gives $q=-1.403$, $m_2=0.9176$, $A_1^*=0.3179$, and
$B_1^*=0.4461$. The corresponding leading coefficients are
$C_0=6.2589$ and $C_{\mathrm{full}}=4.567$. Evaluating
Eqs.~(\ref{eq:r}), (\ref{eq:kappa_bdry}),
(\ref{eq:kappa_coup}), (\ref{eq:kappa_diff}),
(\ref{eq:kappa_dyn_sum}), and
(\ref{eq:eta}) gives $r=0.1750$, $\kappa_{\mathrm{bdry}}=2.874$,
$\kappa_{\mathrm{coup}}=-1.618$, $\kappa_{\mathrm{diff}}=-1.364$,
$\kappa_{\mathrm{dyn}}=-0.1083$,
and $\eta_0=114.5$, yielding $\eta_0r=20.04$ and $C_1=19.93$.

\subsection{Verification of the first-order correction}
\label{app:numerical-verification}

For the correction tests, the analytical zeroth-order optimal
protocol is compared with the numerically optimized finite-time
protocol. Both are evaluated under the full finite-time dynamics.
The fixed zeroth-order protocol is defined by
Eq.~(\ref{eq:app_fixed_potential}); in the quartic parametrization,
its interior controls are
\begin{equation}
 \begin{aligned}
 G_\tau^{[0]}(s)
 &=\tau+\frac{A_1^*-1}{4\mu E_b A_s^*},\\
 F_\tau^{[0]}(s)
 &=\frac{B_1^*}{4\mu E_b A_s^*}.
 \end{aligned}
 \label{eq:app_fixed_quartic_protocol}
\end{equation}
Here the additive $\tau$ follows from
$u_2=-4E_b(g-1)$ and retains the reference quadratic term.
Thus the scaled drift is $b_0+\tau b_r$, with
$b_r(x)=4\mu E_b(x-x^3)$, as required by
Eq.~(\ref{eq:app_fp}). Before and after the operation,
$G_{\mathrm{ref}}=\tau$ and $F_{\mathrm{ref}}=0$.

The work and success probability for these comparisons are evaluated
with the deterministic Fokker--Planck solver. The diagnostic
quantities in Sec.~\ref{sec:114514} are therefore written as
\begin{equation}
 \kappa_{\mathrm{dyn}}^{\mathrm{num}}(\tau)
 =\frac{\tau\langle W\rangle_0^{\mathrm{FP}}-C_0}{\tau},
 \label{eq:app_kappa_dyn_num}
\end{equation}
\begin{equation}
 r^{\mathrm{num}}(\tau)=\frac{p-P_0^{\mathrm{FP}}(\tau)}{\tau},
 \label{eq:app_r_num}
\end{equation}
\begin{equation}
 \kappa_{\mathrm{suc}}^{\mathrm{num}}(\tau)
 =\frac{\tau\langle W\rangle_{\mathrm{opt}}^{\mathrm{FP}}
 -\tau\langle W\rangle_0^{\mathrm{FP}}}{\tau}.
 \label{eq:app_kappa_suc_num}
\end{equation}
The subscript $0$ denotes the fixed analytical protocol in
Eq.~(\ref{eq:app_fixed_quartic_protocol}), whereas $\mathrm{opt}$
denotes the numerically optimized protocol satisfying the target
success-rate constraint. Thus $\langle W\rangle$ in
Eq.~(\ref{eq:kappa_suc_num}) corresponds to
$\langle W\rangle_{\mathrm{opt}}^{\mathrm{FP}}$ here.
Equations~(\ref{eq:app_fixed_work}) and
(\ref{eq:app_optimal_work}) show that their scaled-work difference
is $\tau\eta_0r+O(\tau^2)$, so this comparison with the fully
optimized protocol measures the same first-order constraint-restoration
contribution.

We also compute $\Delta(\tau\langle W\rangle)/\Delta P$, where
$\Delta(\tau\langle W\rangle)
=\tau\langle W\rangle_{\mathrm{opt}}^{\mathrm{FP}}
-\tau\langle W\rangle_0^{\mathrm{FP}}$ and
$\Delta P=p-P_0^{\mathrm{FP}}(\tau)$, and compare it with
$\eta_0=\mathrm{d}C_0/\mathrm{d}p$. As $\tau$ decreases, the
four diagnostics converge to $\kappa_{\mathrm{dyn}}$, $r$,
$\eta_0r$, and $\eta_0$, respectively, as shown in
Fig.~\ref{fig:coefficient_compare}.

% Original bibliography database retained. The unused SM entry is not cited.
\bibliography{references}

\end{document}